\documentclass[longauth]{aa} 
\usepackage[switch]{lineno}  

\usepackage{graphicx}

\usepackage{txfonts}
\usepackage{amsmath}
\usepackage{siunitx}
\usepackage{hyperref}
\usepackage[super]{nth}

\usepackage[T1]{fontenc}
\newcommand\lat{{\it Fermi}-LAT}

\newcommand\xrt{{\it Swift}-XRT}

\usepackage{ulem}
\usepackage{xparse}
\usepackage{xcolor}

\newif\ifrevision
\revisiontrue    
\revisionfalse 

\newcommand{\rev}[1]{%
  \ifrevision
    \textbf{#1}%
  \else
    #1%
  \fi
}

\newif\ifcanc
\canctrue 
\cancfalse

\DeclareRobustCommand{\canc}[1]{%
  \ifcanc
    \textcolor{red}{\sout{#1}}%
  \fi
}

\begin{document}

   \title{MAGIC observations of NGC~4278}
    \subtitle{The first low-luminosity radio galaxy with compact jets detected at TeV energies}

   \date{Received XXX; accepted ZZZ}

%
\author{S.~Abe\inst{1} \and
J.~Abhir\inst{2} \and
V.~A.~Acciari\inst{3} \and
A.~Aguasca-Cabot\inst{4} \and
I.~Agudo\inst{5} \and
T.~Aniello\inst{6} \and
S.~Ansoldi\inst{7,38} \and
L.~A.~Antonelli\inst{6} \and
A.~Arbet Engels\inst{8}$^{\star}$ \and
C.~Arcaro\inst{9}$^{\star}$
% ...
\\
(MAGIC Collaboration)\\
E.~Torresi\inst{46}\and
G.~Migliori\inst{45} \and
P.~Grandi\inst{46}
}
\institute { Japanese MAGIC Group: Department of Physics, Kyoto University, 606-8502 Kyoto, Japan
\and ETH Z\"urich, CH-8093 Z\"urich, Switzerland
\and Institut de F\'isica d'Altes Energies (IFAE), The Barcelona Institute of Science and Technology (BIST), E-08193 Bellaterra (Barcelona), Spain
\and Universitat de Barcelona, ICCUB, IEEC-UB, E-08028 Barcelona, Spain
\and Instituto de Astrof\'isica de Andaluc\'ia-CSIC, Glorieta de la Astronom\'ia s/n, 18008, Granada, Spain
\and National Institute for Astrophysics (INAF), I-00136 Rome, Italy
\and Universit\`a di Udine and INFN Trieste, I-33100 Udine, Italy
\and Max-Planck-Institut f\"ur Physik, D-85748 Garching, Germany
\and Universit\`a di Padova and INFN, I-35131 Padova, Italy
\and Instituto de Astrof\'isica de Canarias and Dpto. de  Astrof\'isica, Universidad de La Laguna, E-38200, La Laguna, Tenerife, Spain
\and Croatian MAGIC Group: University of Zagreb, Faculty of Electrical Engineering and Computing (FER), 10000 Zagreb, Croatia
\and Saha Institute of Nuclear Physics, A CI of Homi Bhabha National Institute, Kolkata 700064, West Bengal, India
\and Centro Brasileiro de Pesquisas F\'isicas (CBPF), 22290-180 URCA, Rio de Janeiro (RJ), Brazil
\and IPARCOS Institute and EMFTEL Department, Universidad Complutense de Madrid, E-28040 Madrid, Spain
\and Japanese MAGIC Group: Institute for Cosmic Ray Research (ICRR), The University of Tokyo, Kashiwa, 277-8582 Chiba, Japan
\and University of Lodz, Faculty of Physics and Applied Informatics, Department of Astrophysics, 90-236 Lodz, Poland
\and Centro de Investigaciones Energ\'eticas, Medioambientales y Tecnol\'ogicas, E-28040 Madrid, Spain
\and Departament de F\'isica, and CERES-IEEC, Universitat Aut\`onoma de Barcelona, E-08193 Bellaterra, Spain
\and Universit\`a di Pisa and INFN Pisa, I-56126 Pisa, Italy
\and INFN MAGIC Group: INFN Sezione di Bari and Dipartimento Interateneo di Fisica dell'Universit\`a e del Politecnico di Bari, I-70125 Bari, Italy
\and Armenian MAGIC Group: A. Alikhanyan National Science Laboratory, 0036 Yerevan, Armenia
\and Department for Physics and Technology, University of Bergen, Norway
\and INFN MAGIC Group: INFN Sezione di Torino and Universit\`a degli Studi di Torino, I-10125 Torino, Italy
\and Croatian MAGIC Group: University of Rijeka, Faculty of Physics, 51000 Rijeka, Croatia
\and Universit\"at W\"urzburg, D-97074 W\"urzburg, Germany
\and Technische Universit\"at Dortmund, D-44221 Dortmund, Germany
\and Japanese MAGIC Group: Physics Program, Graduate School of Advanced Science and Engineering, Hiroshima University, 739-8526 Hiroshima, Japan
\and Armenian MAGIC Group: ICRANet-Armenia, 0019 Yerevan, Armenia
\and Croatian MAGIC Group: Josip Juraj Strossmayer University of Osijek, Department of Physics, 31000 Osijek, Croatia
\and Finnish MAGIC Group: Finnish Centre for Astronomy with ESO, Department of Physics and Astronomy, University of Turku, FI-20014 Turku, Finland
\and University of Geneva, Chemin d'Ecogia 16, CH-1290 Versoix, Switzerland
\and Japanese MAGIC Group: Department of Physics, Tokai University, Hiratsuka, 259-1292 Kanagawa, Japan
\and Inst. for Nucl. Research and Nucl. Energy, Bulgarian Academy of Sciences, BG-1784 Sofia, Bulgaria
\and Finnish MAGIC Group: Space Physics and Astronomy Research Unit, University of Oulu, FI-90014 Oulu, Finland
\and Universit\`a di Siena and INFN Pisa, I-53100 Siena, Italy
\and Japanese MAGIC Group: Institute for Space-Earth Environmental Research and Kobayashi-Maskawa Institute for the Origin of Particles and the Universe, Nagoya University, 464-6801 Nagoya, Japan
\and INFN MAGIC Group: INFN Roma Tor Vergata, I-00133 Roma, Italy
\and also at International Center for Relativistic Astrophysics (ICRA), Rome, Italy
\and also at Como Lake centre for AstroPhysics (CLAP), DiSAT, Universit\`a dell'Insubria, via Valleggio 11, 22100 Como, Italy.
\and also at Port d'Informaci\'o Cient\'ifica (PIC), E-08193 Bellaterra (Barcelona), Spain
\and now at Universit\'e Paris Cit\'e, CNRS, Astroparticule et Cosmologie, F-75013 Paris, France
\and also at Dipartimento di Fisica, Universit\`a di Trieste, I-34127 Trieste, Italy
\and Dipartimento di Fisica, Universit\`a di Roma Tor Vergata, Via della Ricerca Scientifica, 1, Roma I-00133, Italy
\and also at INAF Padova
\and INAF, Istituto di Radioastronomia, via P. Gobetti 101, 40129, Bologna, Italy
\and INAF, Osservatorio di Astrofisica e Scienza dello Spazio, via P. Gobetti 101, 40129 Bologna, Italy
}

 
  \abstract
   {}
   {The Large High Altitude Air Shower Observatory (LHAASO) Collaboration has recently reported the first detection at TeV energies of a low-luminosity radio galaxy, NGC~4278. The aim of this work is to investigate the high-energy properties of NGC~4278 during the flaring and subsequent quasi-quiescent states with the Florian Goebel Major Atmospheric Gamma Imaging Cherenkov (MAGIC) telescopes.}
   {NGC~4278 is located in the field of view of two blazars, 1ES~1215+303 and 1ES~1218+304, previously observed by the MAGIC telescopes. Therefore, we re-analyzed MAGIC observations made between 2010 and 2024 on these sources. We also modeled the broadband spectral energy distribution of the source during and after the flaring state at TeV energies.}
   {We did not detect any statistically significant $\gamma$-ray emission from NGC~4278 with MAGIC. The corresponding upper limits obtained using the entire MAGIC dataset ($F_{{\rm UL, }\, >150\, \mathrm{GeV}}=1.5 \times 10^{-12}\, \mathrm{ph \, s^{-1}\, cm^{-2}}$) are consistent with the LHAASO results. The best-fit models obtained for both emission states suggest that the emitting region is strongly particle-dominated, and an efficient acceleration mechanism has to be in action in order to reach TeV energies. The transition between the flaring and quasi-quiescent state cannot be explained by a simple radiative cooling of the emitting particles. The inferred jet power, of the order of $L_{\rm jet}\sim 10^{42}\, \mathrm{erg\,s^{-1}}$, is dominated by the kinetic component in both states and it is in a good agreement with previous, time-averaged observational estimates, supporting the idea that such high-energy flares might be recurrent. The jet, however, remains too weak to break the host-galaxy confinement.
   }
   {}


   \maketitle
%

\footnotetext{$\star$ Corresponding authors: contact.magic@mpp.mpg.de}


\section{Introduction}
\label{sec: introduction}
Low-luminosity ($L_{1.4\, \mathrm{GHz}} \lesssim 10^{23.5} \, \mathrm{W\, Hz^{-1}}$) jetted active galactic nuclei (AGN) are the most common class of radio-loud AGN in the local universe \citep[$z\lesssim0.1$,][]{Best2012}, yet they remained undetected in the very high-energy (VHE; $E>100\, \mathrm{GeV}$) band. They harbor supermassive black holes (SMBHs) that accrete matter at low, sub-Eddington regimes ($\dot{M}\lesssim0.01 \dot{M}_{\mathrm{Edd.}}$, where $\dot{M}_{\mathrm{Edd.}}$ is the Eddington accretion rate), corresponding to $L_{\mathrm{bol}}/L_{\mathrm{Edd}} \sim 10^{-5}$, where $L_{\mathrm{bol}}$ and $L_{\mathrm{Edd}}$ are the bolometric and Eddington luminosities, respectively. When SMBHs accrete at such low rates, it is commonly believed that they are powered by radiatively inefficient accretion flows \citep[RIAFs,][]{Ichimaru1977,Narayan1995}. 

\rev{Low-luminosity active galactic nuclei (LLAGN) represent a heterogeneous population: while some LLAGN display jet morphologies reminiscent of extended Fanaroff–Riley type I radio galaxies (FR I), the majority appear to be compact, parsec-scale, core-dominated systems. Many of these sources share properties with compact symmetric objects (CSOs) -- characterized by small ($<1\, \mathrm{kpc}$), symmetric radio structures -- or with the more recently identified population of FR~0 radio galaxies, which exhibit prominent radio cores but little or no extended emission on kiloparsec scales \citep{Baldi2018,Baldi2025,Giovannini2023}. \cite{Baldi2018} suggested that radio activity in low-luminosity jetted AGN may be intermittent, possibly driven by disk or accretion instabilities \citep[e.g.,][]{Czerny2009}, with activity cycles too short to allow the jets to develop large-scale radio structures \citep{Sadler2014,Sadler2016}. Alternatively, jets in low-luminosity sources may remain compact because of confinement by a dense nuclear environment that inhibits their expansion, although recent results tend to weaken this scenario \citep{Ubertosi2021}. Jets in these low-power systems are observed to be slower and less efficient compared to those in extended radio galaxies \citep{Baldi2023}, possibly indicating a lower black hole spin and/or a weaker magnetic field strength at the event horizon \citep{Grandi2016}. They also do not exhibit fast, large-amplitude variability on daily timescales \citep{Baldi2018}, consistent with relatively large viewing angles with respect to the line of sight \citep[$\theta \gtrsim 10^\circ$,][]{Baldi2018}. At high energies (HE; $E > 100\, \mathrm{MeV}$), only a handful of compact LLAGN have been detected by Large Area Telescope (LAT) onboard Fermi Gamma-Ray Telescope \protect{\citep{Migliori2016,Principe2020,Lister2020,Grandi2016,Swain2025,Paliya2021}}, and none has been firmly detected above 100 GeV.}

\rev{Therefore,} the first detection of a LLAGN with compact radio jets, NGC~4278, at TeV energies was recently achieved by the Large High Altitude Air Shower Observatory \citep[LHAASO][]{Cao2024a,Cao2024b}. It represents a significant milestone in our knowledge of the physics behind compact radio sources. This discovery raises questions about the efficiency in accelerating particles up to TeV energies in low-luminosity jetted AGN, as well as the radiation mechanisms at play.\canc{In general,} \rev{If the VHE emission is produced in the radio mini-jets}, this detection\canc{ opens} \rev{would open} a new window on the physics of extragalactic jets, giving us the unique opportunity to explore the jets' parameter space in such compact, low-luminosity radio structures.

NGC~4278 is a nearby \citep[$D_L \simeq 16.4 \, \mathrm{Mpc}$,][]{Tonry2001} radio-loud AGN. The central SMBH \citep[$M_{\mathrm{BH}}\simeq 3 \times 10^8 \, M_{\odot}$,][]{Wang2003,Chiaberge2005}, accreting in very sub-Eddington regime \citep[$L_{0.5-8 \, \mathrm{keV}}/L_{\mathrm{Edd.}}\simeq 6 \times 10^{-8}$][]{Pellegrini2012}, is likely powered by a RIAF. NGC~4278 shows a compact two-sided \rev{radio} structure on pc-scales \citep[$\sim 4 \, \mathrm{pc}$,][]{Giroletti2005,Boccardi2025} that appears to consist of individual components constituting the mini-jets, instead of a homogeneous well-collimated flow \citep{Giroletti2005}. An expansion velocity of the mini jets of $\beta=v/c\sim 0.76$ (equivalent to a bulk motion of $\Gamma_{\mathrm{bulk}}\simeq 1.54$) from Very Long Baseline Array (VLBA) multi-epoch observations was measured \citep{Giroletti2005}\canc{, estimating a kinematic age of less than 100 yrs}. From multi-epoch VLBA observations, \cite{Giroletti2005} estimated a jet viewing angle of $2^\circ\lesssim\theta \lesssim4^\circ$. 

The source shows long-term variable emission in all bands \citep{Giroletti2005,Cardullo2009,Younes2010,Pellegrini2012, Hernandez2013,Tremblay2016,Chen2023,Cao2024b, Bronzini2024b}. At TeV energies, the source is detected by the Water Cherenkov Detector Array (WCDA) of LHAASO. The LHAASO data were obtained from March 5\textsuperscript{th}, 2021 (MJD 59278) to October 31\textsuperscript{st}, 2023 (MJD 60249) \citep{Cao2024b}. The source was observed in a period of enhanced $\gamma$-ray activity from August 23\textsuperscript{rd}, 2021 (MJD 59449) to January 10\textsuperscript{th}, 2022 (MJD 59589). The source spectrum in the $1\, \mathrm{TeV}-25 \, \mathrm{TeV}$ is well described by a power-law model, defined as $dN/dE = N_0 \left(E/E_0\right)^{-\Gamma}$, with $E_0=3\, \mathrm{TeV}$, $N_0 = (0.74 \pm 0.10) \times 10^{-13} \, \mathrm{TeV^{-1}\, cm^{-2}\, s^{-1}}$, and $\Gamma= 2.39 \pm 0.17$ \citep{Cao2024b}. This result was recently supported by the simultaneous detection of enhanced activity in the X-ray and GeV bands by the X-ray Telescope (XRT) on board the Neil Gehrles Swift Telescope and {\it Fermi}-LAT, respectively \citep{Bronzini2024b}. These findings suggest that the source was in an overall enhanced emission state. For the remaining of the LHAASO observations, the source is observed in a quasi-quiescent state\rev{, with the source still detected by the WCDA}, with an integrated flux about 5 times lower than the flaring state, and a spectrum described by a power-law with $E_0=3\, \mathrm{TeV}$, $N_0 = (0.16 \pm 0.04) \times 10^{-13} \, \mathrm{TeV^{-1}\, cm^{-2}\, s^{-1}}$, and $\Gamma= 2.71 \pm 0.69$ \citep{Cao2024b}.

NGC~4278 is located close ($\sim 1^\circ$ offset) to two well known TeV-emitting blazars: 1ES~1218+304 \citep{Albert2006,Acciari2009,Acciari2010} and 1ES~1215+303 \citep{Aleksic2012,Aliu2013}. For this reason, NGC~4278 falls into the field of view previously observed around these two sources by the Florian Goebel Major Atmospheric Gamma Imaging Cherenkov (MAGIC) telescopes. We therefore re-analyzed archival MAGIC observations carried out on 1ES~1218+304 and 1ES~1215+303. 

The paper is organized as follows: in Section \ref{sec: observations}, we describe the reduction and analysis of the data from the observations collected by MAGIC, \lat{} and \xrt{}; in Section \ref{sec: SED modeling} we present the spectral energy distribution (SED) modeling of the source; finally, we discuss and summarize our results in Section \ref{sec: conclusions}. Throughout this paper, we assume a flat ${\rm \Lambda CDM}$ cosmology with $\Omega_m = 0.315$, $\Omega_\Lambda = 0.685$, and $H_0 = 67.4 \, \mathrm{km \, s^{-1} Mpc^{-1}}$. 

\section{Observations and data analysis}
\label{sec: observations}

\subsection{MAGIC}
The MAGIC telescopes form a system of two Imaging Atmospheric Cherenkov Telescopes (IACTs) located at the Roque de Los Muchachos on the Canary Island of La Palma \citep[$\mathrm{28.762^{\circ} \, N,\, 17.890^{\circ} \,W}$, 2200 m above sea level,][]{Aleksic2016b}. They collect Cherenkov light produced in atmospheric air showers that are generated by $\gamma$-rays reaching the Earth. MAGIC has a field of view of $3.5^\circ$ and is sensitive to $\gamma$-rays with a primary energy above a few tens of GeV up to tens of TeV. The integral sensitivity for point-like sources above 220\,GeV is ($0.83\pm0.03$)\% of the Crab Nebula flux in 50\,hours of observation, assuming a Crab Nebula-like spectrum. The angular resolution at these energies is $<$~0.07\textdegree, while the energy resolution is 16\%. The performance of the instruments and the details of the data analysis are fully described in \citet{Aleksic2016a}. 

Projected in the sky, NGC~4278 is only $\approx 1^\circ$ away from the two TeV blazars 1ES~1215+303 and 1ES~1218+304, which have been regularly observed by MAGIC \citep{Aleksic2012,Colin2011}. We therefore utilize observations of 1ES~1215+303 and 1ES~1218+304 to analyze $\gamma$-ray events at the nominal position of NGC~4278. We selected moonless nights over a period ranging from 2010 to 2024. All observations were performed in the so-called \emph{wobble} mode~\citep{Fomin1994}, pointing the MAGIC telescopes at different sky positions with an offset of $0.4^{\circ}$ around 1ES~1215+303 and 1ES~1218+304. After quality selection, MAGIC collected a total of $86.98\, {\rm hrs}$, split in $\approx 53 \, {\rm hrs}$ on 1ES~1215+303 and $\approx 34\, {\rm hrs}$ on 1ES~1218+304, with zenith angles ranging from $0^{\circ}$ to $50^{\circ}$. 

The data were analyzed using the MAGIC Analysis and Reconstruction Software \citep[\texttt{MARS},][]{Zanin2013,Aleksic2016b}. The data processing pipeline consists of calibration, image cleaning, and calculation of the second moment of the elliptical-shaped camera images of the telescopes, the so-called \emph{Hillas parameterization}. A Random Forest algorithm based on the reconstruction parameters is used to separate the hadronic background from the $\gamma$-ray signal, as well as to estimate the primary $\gamma$-ray energy. Given the different pointings, the offset of NGC~4278 with respect to the observed positions in the sky varied from $0.59^{\circ}$ to $1.36^{\circ}$, requiring tailored Monte Carlo simulations to estimate the instrument response functions at each offset. Additionally, we masked 1ES~1215+303 and 1ES~1218+304 to estimate the background. 
In order to account for variations in the instrument performance over the years, the data had to be divided into separate analysis periods using adapted Monte Carlo simulations and subsequently stacked.

\begin{figure}
    \centering
 ar-X    \includegraphics[width=\linewidth]{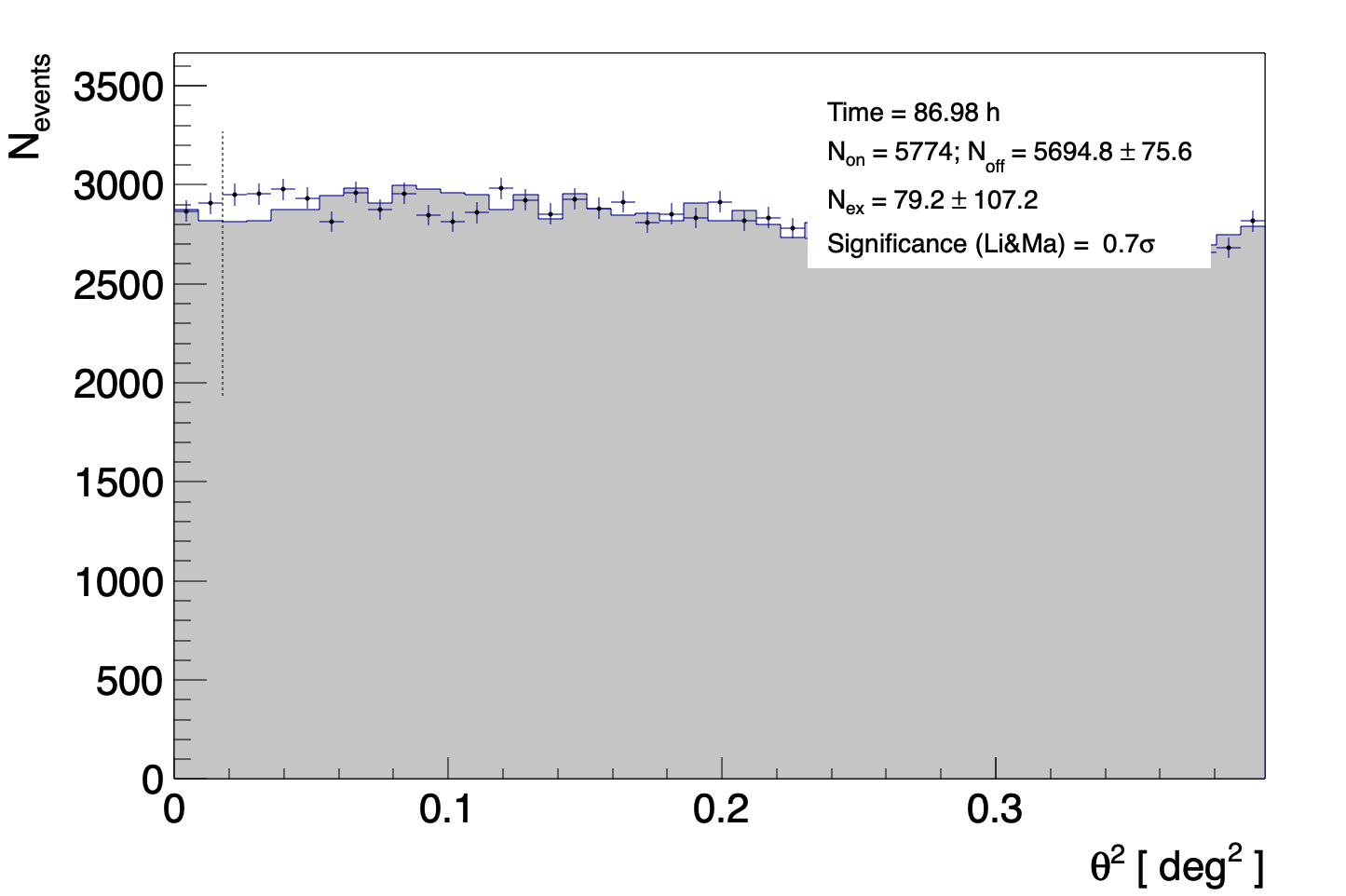}
    \caption{The squared angular distance ($\theta^2$) shows the number of $\gamma$-ray events as a function of the squared angular separation between the reconstructed $\gamma$-ray direction and the position of NGC~4278. The on-source and background signals are shown in black and gray, respectively. The cut in $\theta^2$ used for the calculation of the number of ON and OFF events is shown with the dashed, vertical line.}
    \label{fig: theta_squared plot}
\end{figure}

Within the $\approx 87 \, {\rm hrs}$ of MAGIC observations, no significant $\gamma$-ray signal is detected. Fig. \ref{fig: theta_squared plot} shows the number of detected $\gamma$-ray events as a function of the squared angular distance to the NGC~4278 nominal position for energies above about 100 GeV, while the shaded gray region shows the estimated background. The vertical dashed line defines the boundary of the signal region within which the detection significance is computed. The significance of the $\gamma$-ray signal from the source was computed using the Eq.~17 in \citet{Li1983} and corresponds to $0.7\sigma$. 

\begin{figure}
    \centering
    \includegraphics[width=\linewidth]{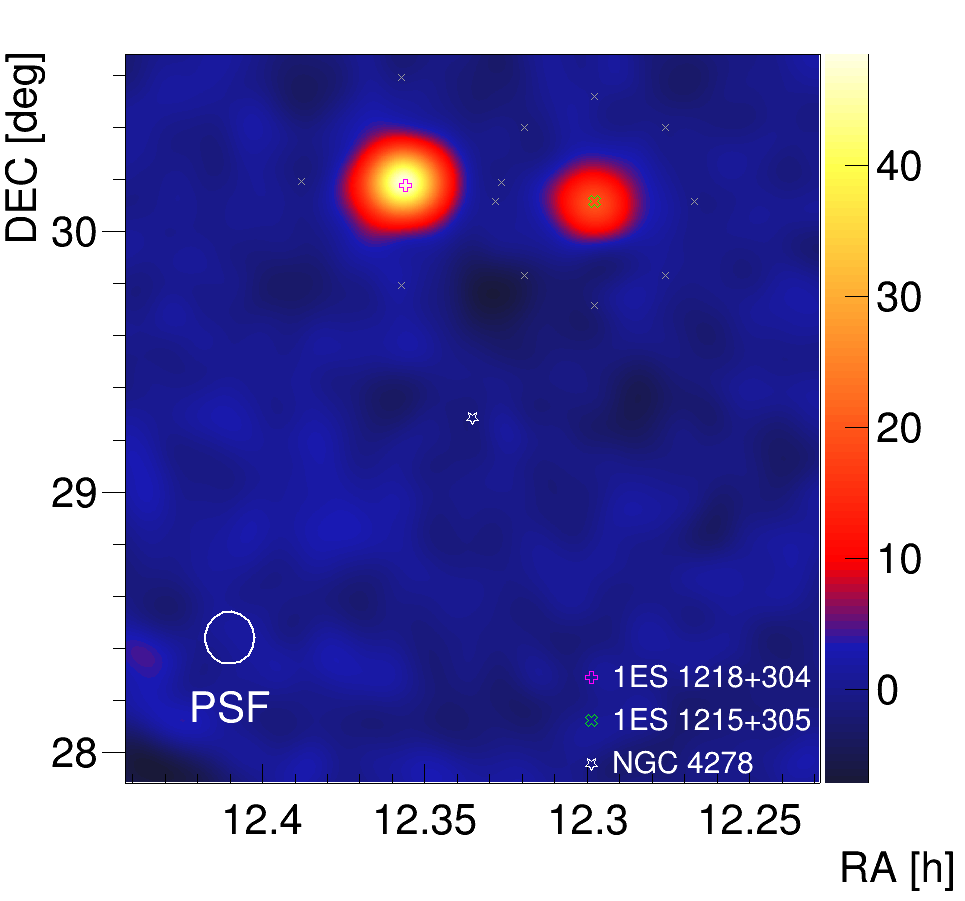}
    \caption{Significance map of the region surrounding NGC~4278. The positions of NGC~4278, 1ES~1218+304, and 1ES~1215+303 are marked with a white star, a pink cross, and a green square, respectively. The colorbar represents the statistical significance in gaussian units. The gray crosses represent the telescopes' pointing positions.}
    
    \label{fig: MAGIC TS map}
\end{figure}

In Fig.~\ref{fig: MAGIC TS map} the significance map above $\approx100$\,GeV of the sky region surrounding NGC~4278 is shown. As for Fig.~\ref{fig: theta_squared plot}, the significance is calculated using the Eq.~17 from \citet{Li1983} and applied on a smoothed and modeled background estimation. While the VHE blazars 1ES~1218+304 and 1ES~1215+303 are clearly detected, no statistically significant signal is found at the position of NGC~4278.

\begin{figure}
    \centering
    \includegraphics[width=\linewidth]{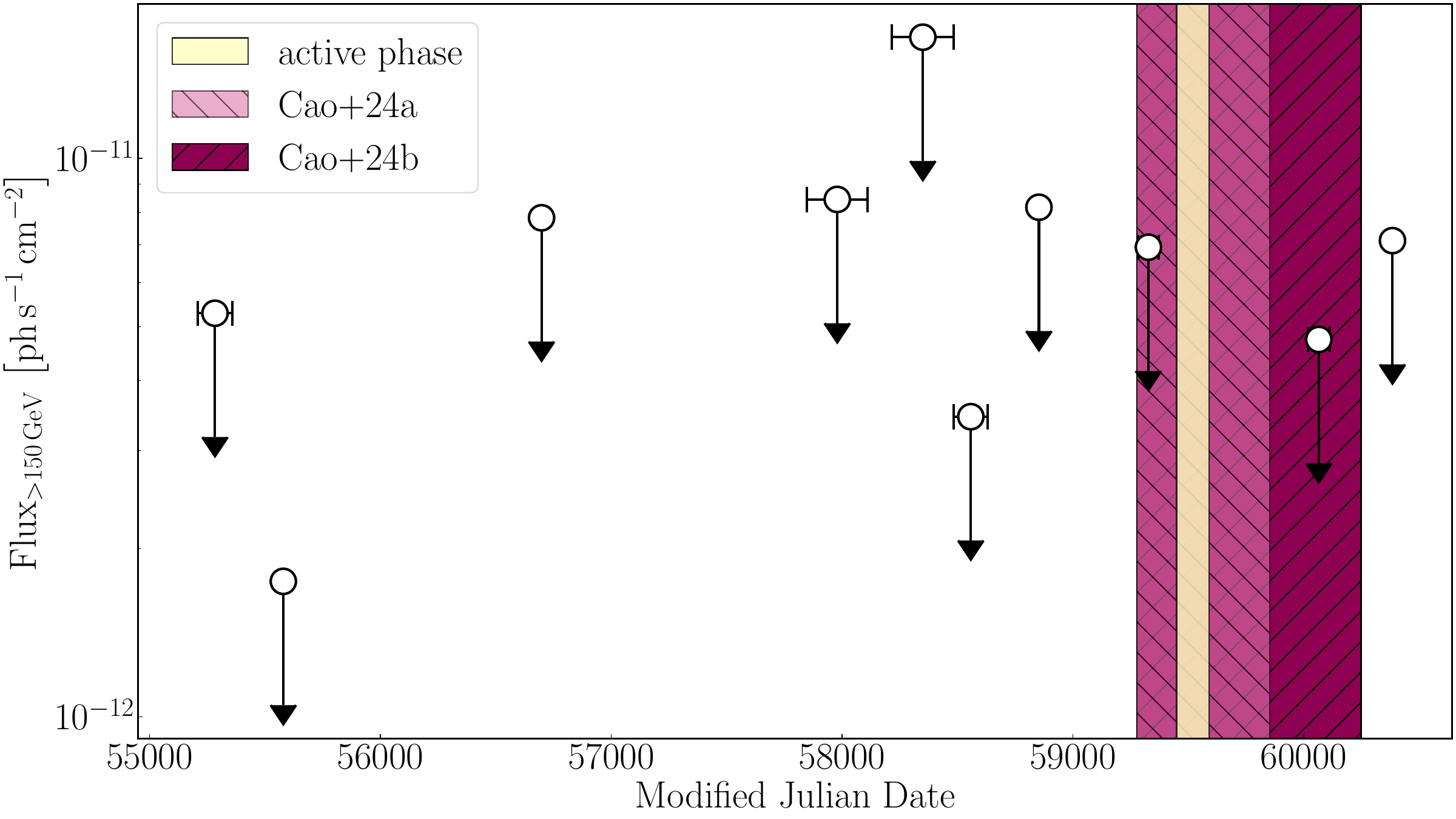}
    \caption{MAGIC light curve above $150\, \mathrm{GeV}$. The 95\% confidence upper limits are shown as black downward arrows. The active phase observed by LHAASO \citep{Cao2024b} is shown as a yellow strip. The period covered in the first LHAASO catalog \citep{Cao2024a} and in the dedicated paper by the LHAASO Collaboration \citep{Cao2024b} are shown as pink and burgundy vertical strips with back diagonal and diagonal hatching lines, respectively.}
    \label{fig: MAGIC lightcurve}
\end{figure}

Due to the absence of a statistically significant detection of a VHE $\gamma$-ray signal from NGC~4278, upper limits (ULs) were derived. The ULs were calculated following \cite{Rolke2005} at a confidence level of 95\%, assuming a power-law index $\Gamma=2$. Fig. \ref{fig: MAGIC lightcurve} shows the yearly-binned MAGIC ULs above $150 \, \mathrm{GeV}$ and the corresponding values are listed in Table \ref{tab: MAGIC lightcurve}. All ULs range from few $10^{-12}$ to few $10^{-11} \,\mathrm{ph \, s^{-1} cm^{-2}}$\canc{, except for the latest data point (few $10^{-10} \,\mathrm{ph \, s^{-1} cm^{-2}}$). This is likely due to a very short integration time (see Table \ref{tab: MAGIC lightcurve})}. We also estimated an average integral flux UL above $150 \, \mathrm{GeV}$ of $F_{{\rm UL, }\; >150\, \mathrm{GeV}}=1.5 \times 10^{-12}\, \mathrm{ph \,s^{-1}\, cm^{-2}}$.

\begin{table}
    \centering
    \caption{MAGIC light curve values.}
    \begin{tabular}{cccc}
    \hline
    \hline
         $t_{\mathrm{start}}$ & $t_{\mathrm{stop}}$ & $t_{\rm eff}$ & 95\% Flux UL \\
         $\mathrm{\left[MJD\right]}$ & $\mathrm{\left[MJD\right]}$ & $\left[\mathrm{hr}\right]$ & $\left[\times 10^{-11} \, \mathrm{ph \, s^{-1} \, cm^{-2}}\right]$\\
         (1) & (2) & (3) & (4)\\
         \hline
         55209.22 & 55357.98 & 23.5 & 0.53\\
         55563.20 & 55596.20 & 21.9 & 0.17\\
         56697.17 & 56699.21 & 3.38 & 0.78\\
         57847.05 & 58111.16 & 4.80 & 0.84\\
         58216.12 & 58483.29 & 3.36 & 1.65\\
         58484.20 & 58631.00 & 14.6 & 0.34\\
         58851.19 & 58854.21 & 3.92 & 0.82\\
         59283.11 & 59370.89 & 4.26 & 0.69\\
         60017.00 & 60113.00 & 9.21 & 0.47\\
         60383.98 & 60384.02 & 1.34 & 0.71\\
    \hline
    \end{tabular}
    \tablefoot{(1-2) Starting and ending date of the observation in Modified Julian Date (MJD), (3) effective integration time, (4) upper limit values at 95\% level of confidence above $150\, \mathrm{GeV}$.}
    \label{tab: MAGIC lightcurve}
\end{table}
The differential upper limits up to $\sim 8 \, \mathrm{TeV}$ for both the post-flare (2023-2024) and the entire (2010-2024) datasets are listed in Table \ref{tab: MAGIC ULs}. In general, our upper limits are consistent with the LHAASO results, constraining the spectrum down to $100 \, \mathrm{GeV}$.

In particular, the majority of the observations ($\sim 88\%$) were\canc{performed} \rev{obtained} before March $5^\text{th}$, 2021 (see Fig. \ref{fig: MAGIC lightcurve}), corresponding to the earliest LHAASO data. During this period, no other observations by other experiments operating in the TeV band are available in the literature. Furthermore, no MAGIC observations were conducted during the flaring state observed by LHAASO (see Fig. \ref{fig: MAGIC lightcurve}).

\begin{table}
    \centering
    \caption{MAGIC differential upper limits.}
    \begin{tabular}{ccc}
    \hline
    \hline
         $E_{\mathrm{ref}}$ & 95\% Flux UL post. & 95\% Flux UL tot.\\
         $\mathrm{TeV}$ &  $\left[\times 10^{-12} \, \mathrm{TeV \, s^{-1} \, cm^{-2}}\right]$ &  $\left[\times 10^{-12} \, \mathrm{TeV \, s^{-1} \, cm^{-2}}\right]$\\
         (1) & (2)\\
         \hline
            0.11 & 1.67 & 5.59 \\
            0.20 & 1.20 & 3.25\\
            0.38 & 0.69 & 1.38\\
            0.70 & 0.53 & 0.87\\
            1.29 & 0.28 & 0.82\\
            2.38 & 0.23 & 0.88\\
            4.39 & 0.26 & 0.86\\
            8.12 & 0.37 & 1.37\\
    \hline
    \end{tabular}
    \tablefoot{(1) Reference energy for the upper limit value, (2-3) flux upper limit at 95\% level of confidence estimated assuming $\Gamma=2$ \rev{from the post-flare (2023-2024) and the entire (2010-2024) datasets}.}
    \label{tab: MAGIC ULs}
\end{table}

\subsection{\lat}

\rev{Being interested in characterizing both the flare and post-flare states of the source,} we inspected the \lat{} \citep{Atwood2009} region around NGC~4278, integrating data from October 1\textsuperscript{st}, 2022 (MJD 59853) to July 1\textsuperscript{st}, 2024 (MJD 60492). A procedure similar to that detailed in \cite{Bronzini2024b} was applied. We considered both front- and back-converted events in the $100 \, \mathrm{MeV}-1\, \mathrm{TeV}$ energy range and selected the \textsc{SOURCE} class events and the \textsc{P8R3\_SOURCE\_V3} instrument response functions \cite[IRFs,][]{Bruel2018}.

The data analysis was performed by exploiting the open-source Python package \texttt{Fermipy v.1.3.1} \citep{Wood2017} with \texttt{Fermitools v.2.2.0}. 
Only good time intervals (\textsc{LAT\_CONFIG==1} and \textsc{DATA\_QUAL>0}) were selected.

Taking advantage of the event partition introduced with \textsc{Pass 8} \citep{Atwood2013}, we separately analyzed the four different event classes, each characterized by its own point spread function (\textsc{PSF}). Each \textsc{PSF} indicates the quality of the reconstructed direction of the event, from the best (\textsc{PSF3}) to the worst (\textsc{PSF0}). Following the procedure outlined in the 4FGL-DR3 catalog \citep{Abdollahi2020,Abdollahi2022}, we further divided the data into 6 different energy bands. The number of bins and the selection of the optimal zenith-angle cut remain consistent throughout each energy interval, while each type of event requires a different pixel size depending on the \textsc{PSF} width \citep[see Table 1 in][]{Abdollahi2022}. In analogy to \cite{Bronzini2024b}, we additionally split the $30\, \mathrm{GeV}  - 1 \, \mathrm{TeV}$ energy bin into the four \textsc{PSF} components.\canc{In total, 21 components were considered.}

A circular region of interest (RoI) of radius $15^{\circ}$ centered on the radio position of NGC~4278 \citep[$\mathrm{R.A.}=185.028439^{\circ}$, $\mathrm{Dec.}=29.280754^\circ$,][]{Ly2004} was chosen. The model used included the most recent template for Galactic interstellar diffuse emission (\texttt{gll\_iem\_v07.fits}), the isotropic background components appropriate for each \textsc{PSF} (\texttt{iso\_P8R3\_SOURCE\_V3\_PSFx\_v1.txt})\footnote{See \url{https://fermi.gsfc.nasa.gov/ssc/data/access/lat/BackgroundModels.html} for further details.}, and the sources listed in the 4FGL-DR4 catalog \citep{Ballet2023}. The energy dispersion correction (\textsc{EDISP\_BINS=-2}) was enabled, except for the isotropic component. The test statistic (TS) was used to estimate the significance of the sources\footnote{The TS corresponds to the logarithmic ratio of the likelihood of a model with the source being at a given position in a grid ($\mathcal{L}_{\mathrm{src}}$) to the likelihood of the model without that source ($\mathcal{L}_{\mathrm{null}}$), $\mathrm{TS}=2\log \left(\mathcal{L}_{\mathrm{src}}/\mathcal{L}_{\mathrm{null}} \right)$ \citep{Mattox1996}.}. Preliminarily, a fast optimization of the entire RoI was performed, deriving the best-fit model of the RoI. As NGC~4278 is not included in the 4FGL-DR4 catalog, we included a point-source at the $\gamma$-ray coordinate position of the target \citep[$\mathrm{R.A.= 184.99^\circ}$, $\mathrm{Dec.= 29.28^\circ}$,][]{Bronzini2024b}, and parametrized it with a power-law model with spectral index fixed to 2. Then, the parameters of both diffuse components along with the spectral parameters of the sources within a radius of $5^\circ$ surrounding the source of interest were set to vary during the fit. Only the normalizations were left free to vary for sources at a distance between $5^\circ$ and $10^\circ$ from the center of the RoI. The spectral parameters of sources at larger distances, as well as those of the faintest ($\mathrm{TS}<4$), were kept fixed. A binned likelihood fit, minimized using \texttt{Minuit} \citep{James1994}, was performed for all\canc{ 21} components separately, and then summed to obtain the total log-likelihood.

The likelihood analysis returned $\mathrm{TS}=2.5$ for NGC~4278.
According to Wilks' theorem \citep{Wilks1938}, the distribution of the TS asymptotically approaches the $\chi^2$ distribution under the null hypothesis $\mathcal{L}_{\mathrm{null}}$ \citep{Cash1979}. The $p$-value we measured, assuming a $\chi^2$ distribution with 1 degree of freedom, i.e., the normalization of the power-law model, is $\text{\textit{p}-value}\simeq 1.1\times10^{-1}$. This corresponds to a detection significance of approximately $\sim 1.2\sigma$ under the Gaussian assumption.

\subsection{{\it Swift}-XRT}

The \xrt{} data taken during the flaring state were analyzed and results are reported in \cite{Bronzini2024b}. Here, we present for the first time three \xrt{} snapshots of the source obtained between late May 2024 and early June 2024\footnote{We did not include {\it Swift}-UVOT data because the infrared-to-UV emission is expected to be dominated by the host galaxy. Disentangling nuclear emission from the host galaxy contribution with angular resolution of UVOT is then challenging.} (MJD 60461, 60464, 60467). The online tool for \xrt{} \citep[see][for details]{Evans2009} was used to reduce data and extract spectra. We used \texttt{Xspec v.12.14.1} \citep{Arnaud1996} for the spectral analysis. For consistency with previous works in the X-ray band, we report here errors at 90\% level of confidence for one parameter of interest. 

In the analysis, we modeled the background components (low-mass X-ray binaries, thermal gas, etc.) that fall within the extraction region of the source spectrum as described in \cite{Bronzini2024b}. The nuclear emission was modeled with an absorbed power law. The obscuring component, which takes into account both the Galactic and intrinsic absorption, was fixed at $N_{\mathrm{H}}=4\times10^{20} \, \mathrm{cm^{-2}}$ \citep{Pellegrini2012}. The best-fit parameters for the nuclear component were a photon index $\Gamma=2.1 \pm 0.4$ and a normalization $\mathrm{norm}_{\mathrm{{PL}}}=\left(1.9 \pm 0.4\right) \times 10^{-4}\, \mathrm{ph\,s^{-1}\, cm^{-2}\, keV^{-1}}$ at $1\, \mathrm{keV}$. This corresponds to an unabsorbed flux $F_{0.5-8 \, \mathrm{keV}} = \left(8\pm2\right) \times 10^{-13} \, \mathrm{erg \, s^{-1} \, cm^{-2}}$.

\section{SED modeling and discussion}
\label{sec: SED modeling}

To investigate the high-energy properties of NGC~4278, we performed a broadband SED modeling of the source in both the flaring and the quasi-quiescent state. The flaring state was defined between MJD 59274 and MJD 59853, as observed by {\it Fermi}-LAT \citep{Bronzini2024b}, overlapping with the LHAASO active phase \citep{Cao2024b}. The remaining period was classified as the quasi-quiescent state, during which the source was still detected at TeV energies, but exhibiting a lower flux.\canc{ All MAGIC observations analyzed in this work, obtained before and after the flaring episode, were included in the quasi-quiescent state in order to use the most stringent MAGIC constraints at VHEs.} \rev{In the SED modeling of the quasi-quiescent state, the upper limits obtained by MAGIC from the post-flare (2023-2024) dataset by MAGIC were used. In this state, the source shows a higher X-ray flux compared to the faintest state observed by \citet{Pellegrini2012}. This suggests that the post-flare state of the source is characterized by a plateau at VHEs, instead of a net decrease of the flux.}

A jet-origin for the TeV flare was suggested based on variability and energetic arguments \citep{Bronzini2024b,Dominguez2025}. Therefore, we tested a leptonic model for the emission of the mini-jet in both states.
The SED modeling was performed using the \texttt{Jet SED modeler and fitting Tool v.1.3.0} \citep[\texttt{JetSeT},][]{Massaro2006,Tramacere2009,Tramacere2011,Tramacere2020}. 
We considered the simplest yet informative scenario of a jet emitting via a synchrotron mechanism and inverse Compton (IC) scattering. In this model, the emission is produced in a jet region of spherical shape and radius $R$. The spherical volume is uniformly filled with a magnetized relativistic plasma. The energy densities of the relativistic leptons and the magnetic field are $U_{\rm e}$ and $U_B$, respectively. The region is moving with a bulk Lorentz factor $\Gamma_\text{bulk}$, and $\theta$ is the angle between the jet axis and the line of sight to the observer. Interacting with the jet's magnetic field, the electrons in the blob radiate via synchrotron emission. Synchrotron radiation provides seed photons for the IC mechanism \citep[synchrotron self-Compton, hereafter SSC, e.g.,][]{Ginzburg1969,Maraschi1992,Bloom1996}. Given the very radiatively inefficient accretion regime of the source, we do not expect intense external photon fields, i.e. infrared photons from a dusty torus or UV photons from the accretion disk. External Compton \citep[EC,][]{Dermer1993,Sikora1994,Celotti2001,Blazejowski2005} scattering is then expected to be subdominant with respect to the SSC component at high energies. Hence, it is neglected in the model. Here, we assume a neutral blob parameterized by $n_\mathrm{p}/n_\mathrm{e} = 1$, where $n_\mathrm{p}$ and $n_\mathrm{e}$ are the densities of cold protons and relativistic electrons, respectively. Because cold protons do not radiate, their properties do not influence the SED. However, they can substantially affect the estimated kinetic power. A lighter jet, corresponding to a lower $n_\mathrm{p}/n_\mathrm{e}$ ratio, would result in a lower value of the jet kinetic power. Therefore, with $n_\mathrm{p}/n_\mathrm{e} = 1$, we get an upper limit on the true kinetic power of the jet.

In both states, we found the best-fit parameters by performing a Markov chain Monte Carlo (MCMC) calculation. This approach is implemented in \texttt{JetSeT} making use of the \texttt{emcee} package \citep{Foreman-Mackey2013}. Independently of the number of parameters of interest, we used 10000 iterations in our MCMC approach using 1000 burn-in steps and a number of walkers equal to 128. We used the \texttt{iMinuit} minimizer \citep{Dembinski2024} to find a set of initial guesses for the MCMC calculation. When possible, we used observations to constrain the model parameter space. Otherwise, the ranges of the model parameters were set in order to be physically meaningful. Then, for each parameter, uniform priors were used in the selected range. The best-fit model is obtained with the \nth{50} quantile of the posterior distribution for each parameter, whose reported errors refer to the \nth{16} and \nth{84} quantiles. Moreover, we added a 20\% systematic uncertainty to all the data used in the fit to take into account uncertainties related to different instruments resolution and a non-perfect simultaneity. Finally, given the proximity of the source, the attenuation of the $\gamma$-ray flux due to the interaction of $\gamma$-ray photons with the extragalactic background light was considered negligible.

The lack of a good multiwavelength coverage and the degeneracies of the model parameters are the major limitation to testing complex models of the broadband SED of NGC~4278. Then, we assumed the simplest functional form to describe the energy distribution of the radiating electrons: a power-law model defined as
\begin{equation}
    n \left(\gamma\right) = K \gamma^{-p}, \quad \gamma_{\min} \leq \gamma \leq \gamma_{\max}
\end{equation}
where $K$ is the normalization, $\gamma_\mathrm{min}$ and $\gamma_\mathrm{max}$ are the minimum and maximum Lorentz factors of the electrons, and $p$ is the spectral index. So, the actual density of the radiating particles, $N$, is the integral of $n\left(\gamma\right)$ from $\gamma_{\min}$ to $\gamma_{\max}$. 

\begin{figure*}
    \centering
    \includegraphics[width=0.49\linewidth]{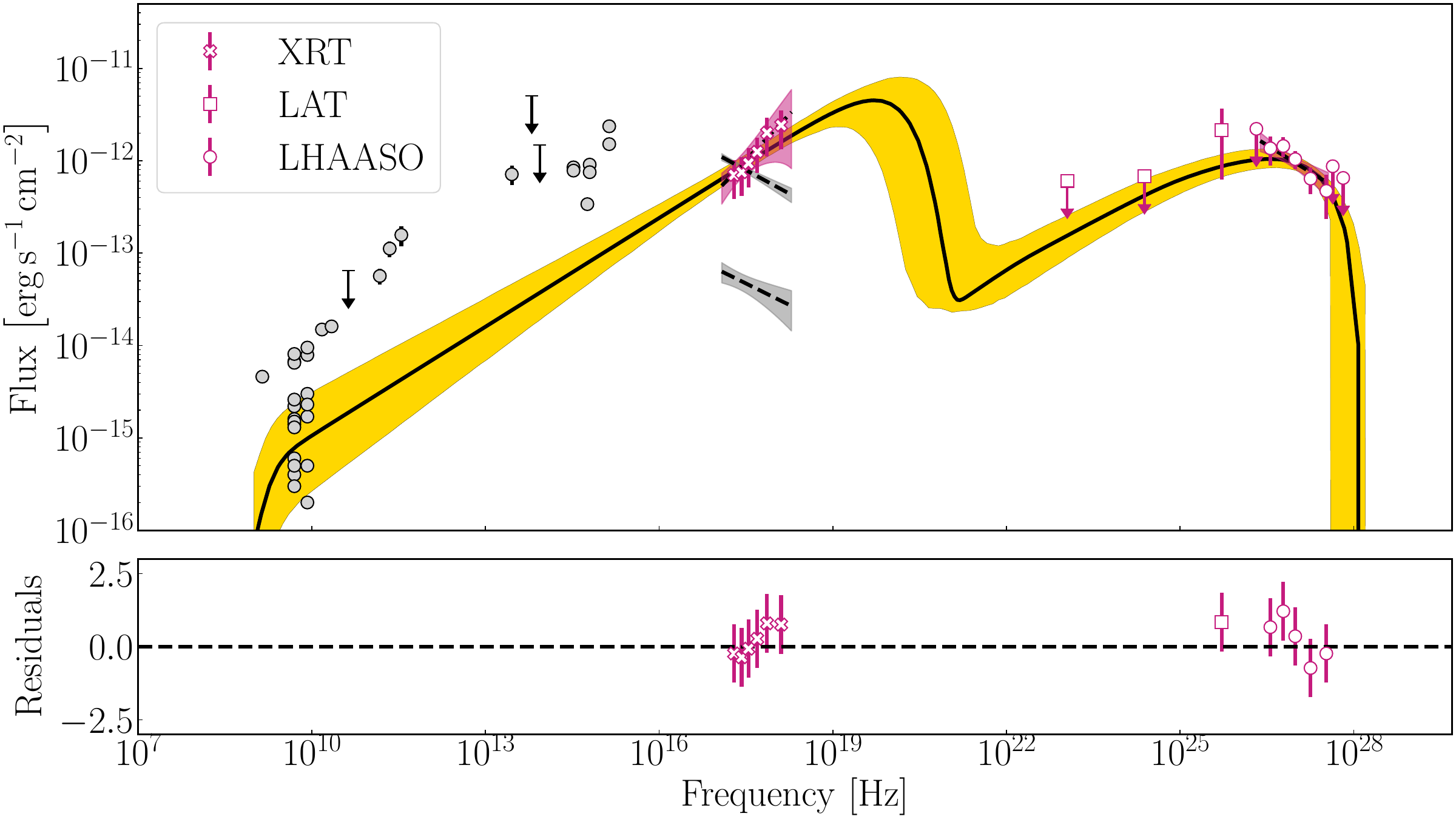}\hfill
    \includegraphics[width=0.49\linewidth]{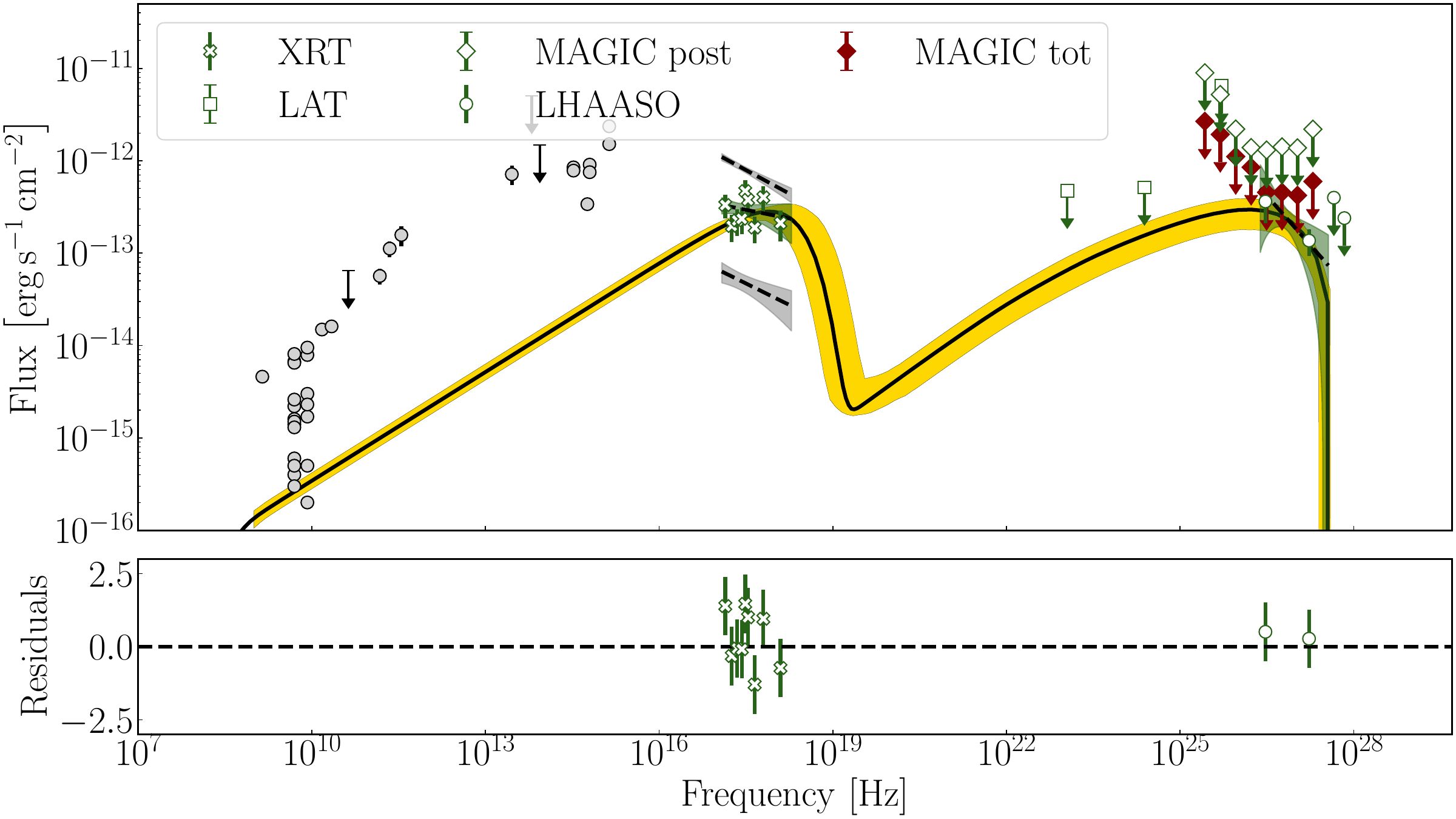}
    \caption{SED modeling of the flaring ({\it left panel}) and the quasi-quiescent ({\it right panel}) state. Data used for the fit are shown in pink and green, respectively, while archival data are shown in gray\rev{ for completeness, but not used in the fitting procedure}. Upper limits are here reported at 95\% level of confidence. \rev{For MAGIC data, we report the upper limits obtained using both the post-flare (2023-2024), in green, and the entire (2010-2024), in burgundy, datasets.} The best-fit models are shown in black and the 16\%-84\% confidence interval of the model is shown as yellow-shaded band.}
    \label{fig: SED model}
\end{figure*}

\subsection{Flaring state}

The multiwavelength high-energy spectrum of the source during the flaring state (see left panel in Figure \ref{fig: SED model}) suggests that, in our SSC framework, the X-ray and GeV-TeV emissions have a different origin: the former is produced via synchrotron radiation, while the latter via SSC scattering. Furthermore, the hardness of the spectrum measured during the flaring state in the X-ray band suggests that the peak of the synchrotron emission is likely to occur at $\nu_{\mathrm{s}}\gtrsim 10^{18}\, \mathrm{Hz}$ \citep{Bronzini2024b}. In this scenario, the spectral index $p$ is directly related to the photon index defined in the X-ray band by $\Gamma=(p+1)/2$ \citep[e.g.,][]{Cerruti2020}. Hence, measuring the photon index in the X-ray band constrains the spectrum of the radiating particles. The $\gamma$-ray component, instead, constrains the peak of the IC component to be at $\nu_{\mathrm{C}} \sim 10^{26-27} \, \mathrm{Hz}$.

The low-energy part of both synchrotron and SSC components is poorly sampled by observations. SED modeling of TeV-emitting BL Lacs typically suggests $\gamma_{\min}\sim10^{2-4}$ \citep[e.g.,][]{Costamante2001, Tavecchio2016}, which can occur as a consequence of electron pre-heating within (mildly relativistic) electron-ion plasma shocks \citep{Arbet2025,Zech2021}. Therefore, we arbitrarily fixed $\gamma_{\min} = 10^2$. Since this choice affects the low-energy tails of both the synchrotron and IC components, it has a negligible impact on the radiated power. However, it has a direct impact on the estimate of the kinetic power of the jet: an increase of $\gamma_{\min}$ implies a decrease of both the electrons' and cold protons' kinetic power.

We also assumed that the bulk motion of the blob and the jet's viewing angle remained the same as measured by \cite{Giroletti2005} ($\theta=3^\circ$, $\beta=0.76$, $\Gamma_{\mathrm{bulk}}=1.54$, hence $\delta=1/\left[\Gamma \left(1-\beta \cos\theta\right)\right]=2.70$).

We set the minimum value of $R$ equal to $10^{15} \, \mathrm{cm}$, based on the typical values obtained from studies of TeV-emitting blazars \citep[e.g.,][]{Celotti2008}. The maximum value was obtained via causality argument from the variability in the VHE band observed by LHAASO \citep{Cao2024b}: with $t_{\mathrm{var}}\sim 14 \, \mathrm{days}$, one gets $R\lesssim 1 \times 10^{17} \, \mathrm{cm}$. In addition, the magnetic field was estimated to be of the order of a few mG \citep{Cao2024b}. The minimum value of the magnetic field was, therefore, set equal to $B_{\min}=1\, \mathrm{mG}$. 
The prior range for the spectral index $p$ was determined from the X-ray observation. The maximum value for the Lorentz factor of the particles can be roughly estimated requiring that the peak of synchrotron emission is at energies $E_s\gtrsim10\, \mathrm{keV}$ \citep[e.g.,][]{Tavecchio1998}: assuming a magnetic field of $B=5\, \mathrm{mG}$, we get $\gamma_{\max}\gtrsim  10^7$.


The best-fit model of the flaring state is shown in the left panel of Figure \ref{fig: SED model}, and the corresponding parameters are listed in Table \ref{tab: sed modeling}. The best-fit for the spectral slope of the radiating electrons is $p=2.2_{-0.1}^{+0.1}$, which is mainly driven by the hard slope observed in X-rays, with the \lat{} data only providing upper limits in the low-energy $\gamma$-ray band. The radius of the emitting region and the magnetic field strength are found to be $R=1.3_{-0.7}^{+2.4} \times 10^{16} \, \mathrm{cm}$ and $B=8_{-5}^{+11} \, \mathrm{mG}$, respectively. The maximum value for the Lorentz factor for the radiating electrons is $\gamma_{\max}=4_{-2}^{+3}\times 10^7$. Internal shocks are efficient mechanisms for accelerating particles up to high values of $\gamma_{\max}\gtrsim10^7$ \citep[e.g.,][]{Bottcher2010, Mimica2012,Vaidya2018,Mukherjee2021}, supporting a synchrotron origin for the X-ray emission. Consequently, the synchrotron peak is at extreme high frequencies, $\nu_{s} \sim 5 \times10^{19} \, \mathrm{Hz}$, and the Compton peak at $\nu_{\rm C} \sim 4 \times10^{26} \, \mathrm{Hz}$, with a low value of the Compton Dominance\footnote{The Compton Dominance ($CD$) parameter is defined as the ratio of the luminosity density at the IC peak to that at the synchrotron peak.} parameter ($CD\sim0.2$). A synchrotron peak at such high frequencies is atypical even for the so-called \emph{extreme high-synchrotron-peaked BL Lac objects} (EHBLs), which generally exhibit synchrotron peaks at $\sim 10^{17\text{--}18}\,\mathrm{Hz}$ \citep{Costamante2001, Costamante2002}. However, it should be noted that the synchrotron peak inferred here is affected by large uncertainties (see Fig.~\ref{fig: SED model}), as it is not directly constrained by the available data. The synchrotron peak frequency depends on both the magnetic field strength and the maximum Lorentz factor of the electrons \citep[$\nu_s \propto \gamma_{\max}^2 B$,][]{Tavecchio1998}. Consequently, the uncertainty on $\nu_s$ propagates directly into large uncertainties on $B$ and $\gamma_{\max}$, which are therefore only weakly constrained by the model. The best-fit value for the magnetic field strength is in a good agreement with the estimated magnetic field $B\gtrsim5\, \mathrm{mG}$, based on variability arguments by \cite{Cao2024b}.

The hard spectral index observed during the flaring state implies that any spectral break must occur above the X-ray band, corresponding to high electron Lorentz factors. However, the absence of observational constraints above $10\, \mathrm{keV}$, together with the reduced sensitivity at TeV energies being in Klein-Nishina \rev{(KN)} regime, prevents a statistical discrimination between a single power-law model and a broken power-law model, such as that proposed by \citet{Lian2024}.

Furthermore, our model suggests \rev{a} strong particle dominance over the magnetic field ($U_B/U_{\mathrm{e}}=2\times 10^{-5}$). Even if this condition is usually required in one-zone SSC models of BL Lac objects, with $U_{B}/U_{\mathrm{e}}\sim10^{-1}-10^{-3}$ \citep[e.g.,][and references therein]{Tavecchio2016,MAGIC2020}, the obtained value is extreme. This strong imbalance can be overcome if the one-zone SSC hypothesis is relaxed: indeed, including also the EC contribution produced within the jet itself, i.e. the spine-layer model \citep{Ghisellini2005}, it is possible to explain the broadband SED of BL Lacs under equipartition conditions \citep[e.g.,][]{Tavecchio2016}. In principle, also the EC scattering against surrounding medium can help in this sense. However, the contribution from the EC scattering is expected to be subdominant due to the likely absence of a standard accretion disk and/or a dusty torus \citep[][and references therein]{Pellegrini2012}.

\begin{table*}
    \centering
    \caption{SED best-fit parameters and estimated jet powers.}
    \begin{tabular}{lcccccc}
    \hline
    \hline \\[-2.5mm]
    \multicolumn{7}{c}{Best-fit parameters}\\
    \hline
    Description & Symbol & Unit & \multicolumn{2}{c}{Flaring state} &  \multicolumn{2}{c}{Quasi-quiescent state}\\
    & & & Value & Prior & Value & Prior\\
    (1) & (2) & (3) & (4) & (5) & (6) & (7)\\
    \hline
    Radius of the emitting region & $R$ & $10^{16}\,\mathrm{cm}$ & $1.3_{-0.7}^{+2.4}$ & $\left[ 0.1,\,10\right]$ & $4.7_{-1.6}^{+2.6}$ & $\left[0.5,\,10\right]$\\
    Magnetic field & $B$ & $\mathrm{mG}$ & $8_{-5}^{+11}$
    & $\left[0.5,\, 30\right]$ & $0.7_{-0.3}^{+0.6}$  & $\left[0.1,\, 20\right]$\\
    Normalization of electrons distribution & $K$ & $ 10^{4}\,\mathrm{cm^{-3}}$ & $1.7_{-1.0}^{+0.9} $ & $\left[0.01,\,3\right]$ & $0.6_{-0.3}^{+0.3}$ & $\left[0.01,\, 1\right]$\\
    Electrons spectral index & $p$ & & $2.2_{-0.1}^{+0.1}$ & $\left[1,\, 3\right]$ & $2.2$ (fixed) & ---\\
    Lorentz factor (max) & $\gamma_{\max}$ & $10^7$ & $4.2_{-1.8}^{+3.2}$ & $\left[1,\, 10\right]$ &$1.8_{-0.5}^{+0.7}$ & $\left[0.1,\, 3\right]$ \\
    Synchrotron-peak frequency & $\nu_s$ & $10^{19}\, \mathrm{Hz}$ & 5 & --- & 0.08 & ---\\
    Compton-peak frequency & $\nu_{\rm C}$ & $10^{26}\, \mathrm{Hz}$ & 3 & --- & 2 & ---\\
    \hline \\[-2.5mm]
    \multicolumn{7}{c}{Energy Budget}\\
    \hline
    Energy-density ratio of magnetic field and electrons & $U_B/U_\mathrm{e}$ & $10^{-4}$& 1 & ---  & 0.03 & --- \\ 
    Radiative power & $L_{\mathrm{rad}}$ & $10^{39} \,\mathrm{erg \, s^{-1}}$ & 8 & --- & 1 & --- \\
    Electrons kinetic power & $L_{\mathrm{e}}$ & $10^{42} \,\mathrm{erg \, s^{-1}}$ & 0.76 & --- & 3.8 & --- \\
    Cold protons kinetic power & $L_{\mathrm{p;\,cold}}$ & $10^{42} \,\mathrm{erg \, s^{-1}}$ & 0.75 & --- & 3.5 & --- \\
    Total kinetic power & $L_{\mathrm{kin}}$ & $10^{42} \,\mathrm{erg \, s^{-1}} $ & 1.5 & --- & 7.3 & --- \\
    Magnetic power & $L_{B}$ & $10^{38} \,\mathrm{erg \, s^{-1}}$ & 1 & --- & 0.1\\
    Total jet power & $L_{\mathrm{jet}}$ & $10^{42} \,\mathrm{erg \, s^{-1}}$ & 1.5 & --- & 7.3 & --- \\
    \hline
    \end{tabular}
    \tablefoot{(1) Description of the input parameter, (2) symbol used for the input parameter, (3) units in which the parameter is expressed, (4–5) best-fit parameter values and variability range for the prior for the jet model during the flaring state, (6–7) best-fit parameter values and variability range for the prior for the jet model during the quasi-quiescent state. Powers are estimated with formulae in \cite{Zdziarski2014}, which are accurate for small bulk motions\rev{, and assuming the best-fit values reported in column (2). Therefore, we highlight that the values of the energy budget reported here should be treated as order of magnitude estimates, given the large uncertainties on the individual best-fit parameters}.}
    \label{tab: sed modeling}
\end{table*}

\subsection{Temporal evolution}
A statistically significant variability in the spectrum normalization was observed by LHAASO at VHE \citep{Cao2024b}. In the $\gamma$-ray band covered by \lat{}\canc{, we cannot observe any statistically significant variation between the period analyzed in this work and the one in due to the poor photon statistics}\rev{, the source is not detected after the flaring state, during which a statistically significant signal was instead detected \citep{Bronzini2024b}}. In the X-ray band, no statistically significant variability at more than $3\sigma$ cannot be claimed. However, \cite{Pellegrini2012} observed a "softening when fading" trend of the source in X-rays. Therefore, it is reasonable to expect this behavior also in this case and we interpreted the new X-ray results as a hint of a such trend.

\begin{figure}
    \centering
    \includegraphics[width=\linewidth]{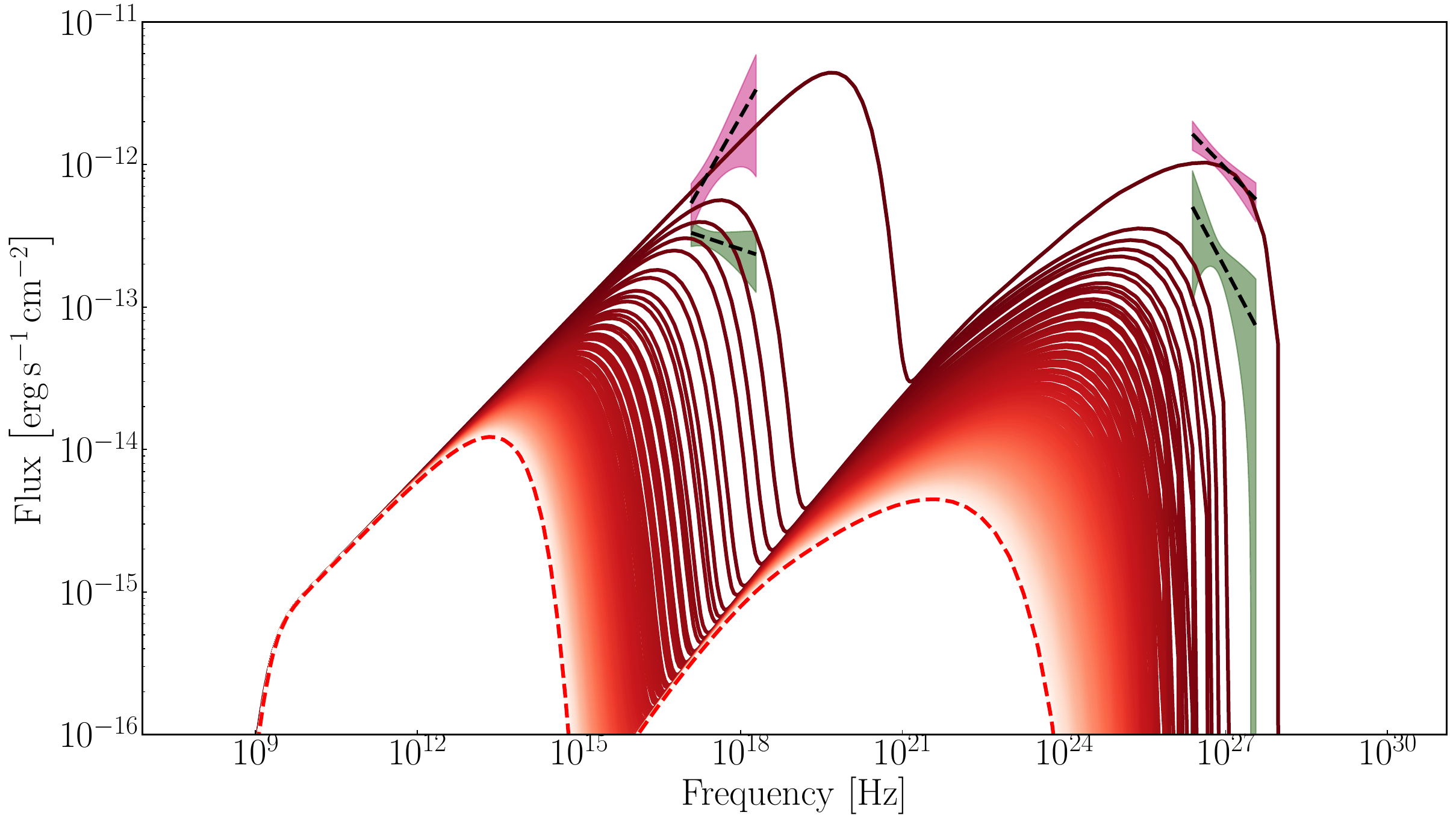}
    \caption{Simulation of the temporal evolution of the blob assuming only radiative cooling in action. The initial parameters for the blob conditions are those of the flaring state, reported in Table \ref{tab: sed modeling}. The duration of the simulation is $t_{\rm sim} = 2\, \mathrm{yr}$ and the time step for clarity reasons is $t_{\rm step}\sim 5\, \mathrm{days}$, both in observer rest-frame. The SED at the end of the simulation is marked as a red dashed line. Here, we show only \xrt{} \citep[][and this work]{Bronzini2024b} and LHAASO \citep{Cao2024b} best-fit models for clarity, with the same color code used in Figure \ref{fig: SED model}.}
    \label{fig: ngc4278 temporal evolution}
\end{figure}

For this reason, we explored the simplest scenario in which the blob that produced the multiwavelength high-energy flare cools radiatively, both via synchrotron and IC cooling. Here, we did not consider any further injection of particles, as well as no particle escape or adiabatic expansion. The initial conditions for the temporal simulation were set equal to the best-fit model obtained in the flaring state and the duration of the simulation is $\sim 2 \,\mathrm{yr}$ in the observed rest-frame. The expected temporal evolution of the SED is shown in Figure \ref{fig: ngc4278 temporal evolution}. As shown in the plot, the fluxes expected after two years in both the X-ray and TeV bands are several orders of magnitude lower than those observed by \xrt{} and LHAASO during the quasi-quiescent state. In particular, from this simulation, one should expect to see the observed fluxes in both the X- and $\gamma$-ray bands in the first 10 days. So, the predicted cooling times are too rapid to reproduce the observed data.

This evidence suggests that a simple blob cooling radiatively cannot link the flaring and quasi-quiescent states in a self-consistent framework. Additional mechanisms and/or different scenarios have to be invoked to explain the observed variability timescales, e.g. further and/or continuous injection. Given these results, we proceed by directly modeling the quasi-quiescent state to constrain the properties of the emitting region during the low state.

\subsection{Quasi-quiescent state}
Since our simulation cannot self-consistently connect the high and quasi-quiescent states, we modeled the emitting region during the quasi-quiescent state using a one-zone SSC framework. Assuming again a simple power-law model for the radiating electrons, the nearly flat spectrum observed by \textit{Swift}-XRT suggests that the synchrotron peak lies in the X-rays. This requires a shift of the synchrotron peak $\nu_s$ at lower frequency with respect to the flaring state. As discussed above, the peak frequency of the synchrotron emission $\nu_s$ is directly related to $\gamma_{\max}$ and $B$, thus, during the quasi-quiescent state, a decrease of the product $\gamma_{\max}^2B$ is expected. Consequently, the ranges for $\gamma_{\max}$ and $B$ were adjusted to also accommodate lower values for each parameter. In particular, a moderate decrease in $\gamma_{\max}$ during the quasi-quiescent state, although not substantial, is expected, as the source continues to emit at TeV energies. In the absence of observational constraints on the spectral index of the radiating particles, we fixed it to the value derived from the flaring state model ($p = 2.2$). The best-fit model of the quasi-quiescent state is shown in the right panel of Figure \ref{fig: SED model}, and its parameters are listed in Table \ref{tab: sed modeling}.

Comparing the high and the quasi-quiescent states, statistically significant differences in the model parameters cannot be claimed, given the large degeneracies in the model parameters and the sparse datasets. \rev{The two states cannot be firmly distinguished and the transition from the flaring state to the post-flare plateau could be caused by a temporary variation in certain parameters.}\canc{ However, w} \rev{W}ith respect to the flaring state, a one order of magnitude lower value of the magnetic field strength was hinted during the quasi-quiescent state ($B=0.7_{-0.3}^{+0.6}\, \mathrm{mG}$), with also a hint of a larger value of the radius of the emitting region by a factor of 3 ($R=4.7_{-1.6}^{+2.6} \times 10^{16} \, \mathrm{cm}$). As expected, during the quasi-quiescent state an efficient acceleration mechanism was still in action to keep $\gamma_{\max}\sim 10^7$. While the peak of the IC component was still at $\nu_{\rm C}\sim10^{26} \, \mathrm{Hz}$, the peak of the synchrotron component shifted at $\nu_s\sim8\times 10^{17} \, \mathrm{Hz}$. This behavior can be explained if one considers that the highest energy electrons ($\gamma\gtrsim 10^7$) are in KN regime, thus they do not contribute to the SSC component significantly. Even if a such value of the synchrotron peak frequency would led NGC~4278 to be still classified as an EHBL, the shift of $\sim 2$ orders of magnitude in the frequency from the high value measured during the flaring state suggests that NGC~4278 might behave as a transitional BL Lac object. In this regard, the soft\rev{er} X-ray spectral \rev{indices} measured by \citet{Pellegrini2012} suggest\canc{s} that the synchrotron peak can further shift at lower frequencies in lower states. 
This behavior is observed in other well-known TeV blazars, i.e. Mkn 501 and 1ES~2344+514 \citep[e.g.,][and references therein]{Pian1998,Ahnen2018,MAGIC2024, 2024A&A...682A.114M}. As in the case of the flaring state, during the quasi-quiescent state a strong particle dominance over the magnetic field ($U_B/U_\mathrm{e} \sim 10^{-6}$) is required.

However, given the\canc{large uncertainties} \rev{numerous caveats} (i.e. non-simultaneity of data, different extraction regions, different integration times), these considerations cannot be asserted with statistical significance, and they should be treated with caution.

\subsection{Jet powers}

To estimate the jet power of the source, we used the formulae in \cite{Zdziarski2014}, which are accurate for slow bulk motions ($\Gamma_{\mathrm{bulk}}\lesssim 5$). In particular, assuming a pure power-law pair plasma, the kinetic power in radiating electrons is given by
\begin{equation}
    L_{\mathrm{e}} = \pi \eta N \langle \gamma \rangle m_{\mathrm{e}} c^3 \beta \left(\Gamma_{\mathrm{bulk}} R \right)^2,
    \label{eq: electrons power}
\end{equation}
where $\eta$ is the average adiabatic index, whose value ranges from 4/3 to 5/3, $\langle \gamma \rangle$ is the mean value of the Lorentz factor of the radiating particles in the comoving frame, $m_{\mathrm{e}}$ is the electron mass, and $c$ is the speed of light in vacuum. Here, we assume $\eta=4/3$.
The kinetic power in cold protons is, instead, estimated as
\begin{equation}
    L_{\mathrm{p}} = \pi \eta N_{\mathrm{p}} m_{\mathrm{p}} c^3 \beta \Gamma_{\mathrm{bulk}}\left(\Gamma_{\mathrm{bulk}} -1\right) R^2,
    \label{eq: protons power}
\end{equation}
where $N_{\mathrm{p}}$ is the proton density, and $m_{\mathrm{p}}$ is the proton mass. 
The power in the magnetic field obtained from
\begin{equation}
    L_{B} = \eta_{B} \frac{B^2}{8}c\beta \left(\Gamma_{\mathrm{bulk}} R \right)^2,
    \label{eq: magnetic power}
\end{equation}
where $\eta_{B}$ is the magnetic adiabatic index, which value ranges from 4/3 to 2, according to magnetic field properties \citep{Leahy1991}. Here, we assume $\eta_{B}=4/3$, corresponding to a fully tangled magnetic field.
The total jet power, in the blob, is then
\begin{equation}
    L_{\mathrm{jet}} = L_{\mathrm{e}} +L_{\mathrm{p}}+L_{B}.
\end{equation}

\noindent The estimated jet powers are listed in Table \ref{tab: sed modeling}\rev{ and are calculated assuming the best-fit value for each model parameter. However, given the large uncertainties on the individual parameters, these values should be treated as order-of-magnitude estimates of the energy budget of the jet.}

The jet power is comparable in both states and of the order of $L_{\rm jet}\sim 10^{42}\,\mathrm{erg\,s^{-1}}$. More than $99\%$ of this power is stored in kinetic form, roughly equally shared between relativistic electrons and cold protons. Only a very small fraction ($\lesssim 1\%$) is radiated away, indicating that most of the jet power is used instead to inflate the radio structure and/or released into the interstellar medium of the host galaxy.

Remarkably, these estimates of the jet power are in excellent agreement with the observational constraints reported by \citet{Pellegrini2012}. Using both the $5\,\mathrm{GHz}$ radio core luminosity \citep{Merloni2007} and the cavity power \citep{Cavagnolo2010}, \citet{Pellegrini2012} inferred a jet power for NGC~4278 of $L_{\rm jet}\sim (1\text{--}2)\times10^{42}\,\mathrm{erg\,s^{-1}}$.

This agreement also supports our initial choice of $\gamma_{\min}=10^{2}$, which primarily affects the kinetic power budget. Observationally, as discussed above, $\gamma_{\min}$ cannot be constrained with the available data and must therefore be assumed. Adopting a lower value, $\gamma_{\min}\sim 1\text{--}10$, increases the kinetic power carried by cold protons by one to two orders of magnitude, yielding jet powers as high as $L_{\rm jet}\sim10^{44}\,\mathrm{erg\,s^{-1}}$. Since the jet power estimated by \cite{Pellegrini2012} represents a time-averaged value, such high powers may be acceptable during flaring episodes. However, they are difficult to reconcile with the quasi-quiescent state, which is expected to be relatively steady and therefore more consistent with the average jet power. Conversely, assuming a larger value, such as $\gamma_{\min}\sim10^{3}$, slightly reduces the total jet power, still of the order of $L_{\rm jet}\sim 10^{42} \, \mathrm{erg\, s^{-1}}$, which in this case becomes entirely dominated by the kinetic energy of the electrons.

For lower-luminosity sources, both the radio duty cycle and the jet lifetimes are more uncertain, and the activity appears to be more frequently retriggered \citep{Best2005}. In this context, \citet{Pellegrini2012} suggested that NGC~4278 may represent an extreme case of frequent triggering of low-power jet components. The SED modeling presented here further supports this scenario: episodic flaring events such as the one detected by \citet{Bronzini2024b,Cao2024b}, can account for the jet power observationally inferred by \citet{Pellegrini2012}.

This interpretation is also consistent with the jet structural and dynamical properties reported by \citet{Giroletti2005}, which indicate jets composed of individual components rather than a homogeneous, continuous flow. Moreover, it is in good agreement with numerical simulations, which suggest that, in modest power jets ($L_{\rm jet}\lesssim10^{43}\, \mathrm{erg\,s^{-1}}$) such flaring episodes may not be sufficiently energetic to escape confinement by the host galaxy \citep{Mukherjee2017}.

\section{Conclusions}
\label{sec: conclusions}

In this work, we present the study of the high-energy properties of NGC~4278, the first low-luminosity AGN with compact jets detected at TeV energies by LHAASO \citep{Cao2024a,Cao2024b}.

\rev{To this end, we re-analyzed archival MAGIC observations obtained between 2010 and 2024, during which NGC~4278 was serendipitously observed while targeting the nearby blazars 1ES~1218+304 and 1ES~1215+303. We did not obtain any detection; however, this time interval does not include the flare observed by LHAASO \citep{Cao2024b}. We therefore derived upper limits that are consistent with the LHAASO results.}

Since the source lies within the field of view of the two blazars, we took advantage of the serendipitous MAGIC observations of NGC 4278 collected over the 2010–2024 period. We did not obtain any detection; however, this time interval does not include the flare observed by LHAASO. We therefore derived upper limits that are consistent with the LHAASO results.

We modeled the broadband SED in a simple one-zone SSC framework assuming a jet origin for the high-energy emission: our best-fit results suggest that the high-energy emission is produced by a compact, particle-dominated region in the mini-jets during the TeV flare. Starting from the best-fit model obtained during the enhanced TeV state, we explored the evolution of this electron population, considering only radiative losses (both via synchrotron radiation and IC scattering). We showed that the cooling time is too rapid and that it cannot fit the observations after the flare both in X- and $\gamma$-rays. Additional mechanisms, i.e. additional injection episodes and/or dynamical evolution of the blob along the jet, need to be considered. We fitted the broadband SED of the source using data up to 2 years after the flare. The modeling presented here supports the idea that NGC~4278 can be interpreted as an extreme case of recurrent low-power jet launching. These episodic flaring events are sufficient to reproduce the time-averaged jet power inferred from observations. This picture is consistent with VLBI evidence for a jet composed of discrete components \citep{Giroletti2005} and with numerical simulations indicating that such events are insufficiently energetic to escape host galaxy confinement \citep{Mukherjee2017}. Even if the inferred jet power is modest, it should be noted that the mini-jets in NGC~4278 can efficiently accelerate particles up to $\gamma_{\max}\gtrsim10^7$.

Our results break new ground for further future investigations. Low-luminosity AGN with compact radio jets may be a promising new class of sources for the upcoming Cherenkov Telescope Array Observatory \citep{CTA2019,Hofmann2023}.

\begin{acknowledgements}
Author's contribution: A. Arbet-Engels: MAGIC analysis,  paper reviewing; C. Arcaro: MAGIC analysis and \texttt{gammapy} cross-check, paper reviewing; E. Bronzini: project management and coordination, \xrt{} and \lat{} data analysis, SED fitting, discussion and interpretation, paper drafting; the rest of the authors have contributed in one or several of the following ways: design, construction, maintenance and operation of the instrument(s) used to acquire the data; preparation and/or evaluation of the observation proposals; data acquisition, processing, calibration and/or reduction; production of analysis tools and/or related Monte Carlo simulations; overall discussions about the contents of the draft, as well as related refinements in the descriptions.

We would like to thank the Instituto de Astrof\'{\i}sica de Canarias for the excellent working conditions at the Observatorio del Roque de los Muchachos in La Palma. The financial support of the German BMFTR, MPG and HGF; the Italian INFN and INAF; the Swiss National Fund SNF; the grants PID2022-136828NB-C41, PID2022-137810NB-C22, PID2022-138172NB-C41, PID2022-138172NB-C42, PID2022-138172NB-C43, PID2022-139117NB-C41, PID2022-139117NB-C42, PID2022-139117NB-C43, PID2022-139117NB-C44, CNS2023-144504 funded by the Spanish MCIN/AEI/ 10.13039/501100011033 and "ERDF A way of making Europe"; the Indian Department of Atomic Energy; the Japanese ICRR, the University of Tokyo, JSPS, and MEXT; the Bulgarian Ministry of Education and Science, National RI Roadmap Project DO1-400/18.12.2020 and the Academy of Finland grant nr. 320045 is gratefully acknowledged. This work has also been supported by Centros de Excelencia ``Severo Ochoa'' y Unidades ``Mar\'{\i}a de Maeztu'' program of the Spanish MCIN/AEI/ 10.13039/501100011033 (CEX2019-000918-M, CEX2021-001131-S, CEX2024001442-S), by AST22\_00001\_9 with funding from NextGenerationEU funds and by the CERCA institution and grants 2021SGR00426, 2021SGR00607 and 2021SGR00773 of the Generalitat de Catalunya; by the Croatian Science Foundation (HrZZ) Project IP-2022-10-4595 and the University of Rijeka Project uniri-prirod-18-48; by the Deutsche Forschungsgemeinschaft (SFB1491) and by the Lamarr-Institute for Machine Learning and Artificial Intelligence; by the Polish Ministry of Science and Higher Education grant No. 2025/WK/04; by the European Union (ERC, MicroStars, 101076533); and by the Brazilian MCTIC, the CNPq Productivity Grant 309053/2022-6 and FAPERJ Grants E-26/200.532/2023 and E-26/211.342/2021.

The \textit{Fermi} LAT Collaboration acknowledges generous ongoing support
from a number of agencies and institutes that have supported both the
development and the operation of the LAT as well as scientific data analysis.
These include the National Aeronautics and Space Administration and the
Department of Energy in the United States, the Commissariat \`a l'Energie Atomique
and the Centre National de la Recherche Scientifique / Institut National de Physique
Nucl\'eaire et de Physique des Particules in France, the Agenzia Spaziale Italiana
and the Istituto Nazionale di Fisica Nucleare in Italy, the Ministry of Education,
Culture, Sports, Science and Technology (MEXT), High Energy Accelerator Research
Organization (KEK) and Japan Aerospace Exploration Agency (JAXA) in Japan, and
the K.~A.~Wallenberg Foundation, the Swedish Research Council and the
Swedish National Space Board in Sweden.

Additional support for science analysis during the operations phase is gratefully 
acknowledged from the Istituto Nazionale di Astrofisica in Italy and the Centre 
National d'\'Etudes Spatiales in France. This work performed in part under DOE 
Contract DE-AC02-76SF00515.

The research activity reported in this paper was carried out with contribution of the Next Generation EU funds within the National Recovery and Resilience Plan (PNRR), Mission 4 - Education and Research, Component 2 - From Research to Business (M4C2), Investment Line 3.1 - Strengthening and creation of Research Infrastructures, Project IR0000012 – CTA+ - Cherenkov Telescope Array Plus''.

E.B. acknowledges ``FERMI LAT - Accordo ASI n. 2023-17-HH." for financial support.
\end{acknowledgements}

\bibliographystyle{aa} 
\bibliography{bibliography}

\end{document}